\documentstyle[vsolj01,graphicx,natbib]{article}

\RequirePackage[T1]{fontenc}

\def\cite{\citealt}

\begin{document}

\title{On the superhumps and mass ratio of CzeV404}

\author{Taichi Kato$^1$}
\author{$^1$ Department of Astronomy, Kyoto University,
       Sakyo-ku, Kyoto 606-8502, Japan}
\email{tkato@kusastro.kyoto-u.ac.jp}

\begin{abstract}
CzeV404 is an SU UMa-type dwarf nova in the period gap.
\citet{kar21czev404} (arXiv:2107.02664) recently
published photometric and spectroscopic observations
and obtained a mass ratio $q$=0.16, which is in severe
disagreement of $q \sim$0.32 estimated from superhump
observations \citep{bak14czev404}.
I here present what analysis was wrong or outdated
in \citet{bak14czev404} and provide a new value of
$q$=0.247(5), consistent with the known behavior
of superhumps and the evolution of cataclysmic variables.
CzeV404 does not look like an unusual dwarf nova
as suggested by \citet{kar21czev404} and I discuss
that the link between SW Sex and SU UMa systems
suggested by \citet{kar21czev404} is not supported.
\end{abstract}

   CzeV404 is an SU UMa-type dwarf nova in the period gap
\citep{bak14czev404}.  \citet{kar21czev404} recently
published photometric and spectroscopic observations
and obtained a mass ratio $q$=0.16, which is in severe
disagreement of $q \sim$0.32 estimated from superhump
observations \citep{bak14czev404}.

   Here I re-examined the superhump observations in
\citet{bak14czev404} and solved the discrepancy.
\citet{bak14czev404} used all the superhumps timing
observations to derive a strongly negative
$P_{\rm dot} = \dot{P}/P$.  Such interpretations were
common (e.g. \cite{nog03var73dra}; \cite{kat03v877arakktelpucma};
\cite{ole03ksuma}; \cite{rut07v419lyr}) before the establishment of
the superhumps stages (A, B and C) in \citet{Pdot}
but are now outdated.
In \citet{Pdot}, it was shown that all previously
reported values of large negative $P_{\rm dot}$ simply
reflected the stage transitions of superhumps rather
than smooth, monotonous period variations.

   CzeV404 \citep{bak14czev404} was a textbook case of
stage A-B transition.  Their observations
started on the rising branch of the superoutburst
(HJD 2456856) and the earliest phase of the superhump
development (stage A) was caught up to HJD 2456859.
Looking at their light curve (in their figure 3),
it is difficult to recognize superhumps on the first
night (HJD 856), but growing superhumps were evidently
caught on the second night (HJD 857).
It is likely that their superhump detections for $E$=0, 1
on HJD 856 were spurious.  Using their superhump maxima
for 10$\le E \le$20, I obtained a period of 0.1060(1)~d.
This value corresponds to
$\epsilon^* \equiv 1-P_{\rm orb}/P_{\rm SH}$, where
$P_{\rm orb}$ and $P_{\rm SH}$ represent orbital and
superhump periods, respectively,
of 0.075(1).  Using the relation between $\epsilon^*$
and $q$ in \citet{kat13qfromstageA}, this value
corresponds to $q$=0.247(5), which is a very reasonable
value for a cataclysmic variable in the period gap
(see e.g. \cite{kni11CVdonor}).

   The problems in the analysis by \citet{bak14czev404}
were: (1) they used the entire phases of superoutbursts,
without paying attention to superhump stages and
(2) they used an inappropriate formula to estimate $q$,
which is only applicable to stage B superhumps
or novalike stars [but this formula does not consider
the effect of the pressure effect and should be considered
as an experimental formula; see \citet{kat13qfromstageA}
for detailed discussions].
The statement in \citet{bak14czev404}:
``The tidal instability (Whitehurst 1988) of the disk
starts to work effectively for binaries with a mass ratio
$q$ below 0.25. This assumption was used by Osaki (1989) in
the TTI (thermal-tidal instability) model to explain
the phenomenon of superoutburstsand superhumps.
That is why such a high value of $q$ in CzeV404 poses
a serious problem for thesuperhump mechanism.'' is
now unsubstantiated.
I must add, however, there is indeed a case of
$q$=0.31--0.34 object, BO Cet, which showed a superoutburst
and superhumps \citep{kat21bocet}.  $q$=0.25 is not
an absolute limit.

   Just for completeness, large negative $P_{\rm dot}$
of MN Dra mentioned in \citet{bak14czev404} was a result
of stage A-B transition as clarified by \citet{kat16v1006cyg}
and \citet{Pdot8}.

   The resultant $q$=0.247(5) for CzeV404 does not match
$q$=0.16 from eclipse modeling in \citet{kar21czev404}.
There must have been something wrong with the treatment
in \citet{kar21czev404} (for example, they assumed
that the disk extends to the tidal truncation radius,
which is clearly an overestimate).  Eclipses analysis
needs to be redone with better quality data.
The statement in \citet{kar21czev404}:
``This is lower than the value estimated from the superhump
period applying different $\epsilon \sim q$ relationships.
However, we note that a similar discrepancy is also observed
in some other eclipsing systems (Kato \& Osaki 2013)''
is apparently a misunderstanding.  There is no difference
in the $\epsilon \sim q$ relationship between eclipsing
and non-eclipsing systems.

   Spectroscopy is outside the scope of this paper
and I only make short comments.  The spectroscopic results
by \citet{kar21czev404} needs to be treated with caution:
(1) It was not clearly described whether they considered
Jacobian in transforming the Doppler tomogram
in the velocity space to real coordinates in obtaining
their figure 12.  If this was not considered, the resultant
map gives an incorrect impression.
(2) SW Sex phenomenon is generally considered to be seen
in systems with high mass-transfer rates (novalike systems
above the thermal stability) while
the observations by \citet{kar21czev404} were made
in lower mass-transfer states.  We showed that BO Cet,
which had been considered as an SW Sex star
\citep{rod07newswsex}, had an orbital light curve
not resembling that of an SW Sex star \citep{kat21bocet}.
The mass of the white dwarf in BO Cet was estimated to be
larger than 1.0$M_\odot$.  If the mass of the white dwarf
in CzeV404 is indeed high ($\sim$1.0$M_\odot$,
\cite{kar21czev404}), the features resembling those of
SW Sex stars in CzeV404 may be a result of
a massive white dwarf as in BO Cet.

\end{document}